\documentclass[%
 reprint,
 amsmath,amssymb,
 aps,
]{revtex4-1}
\usepackage[final]{graphicx}
\usepackage{bm}
\usepackage{afterpage}
\usepackage{color}
\usepackage{comment}
\usepackage[colorlinks=true,urlcolor=blue,linkcolor=blue,citecolor=blue]{hyperref}

\emergencystretch=\maxdimen
\begin{document}

\title{Unconventional pairing symmetry of interacting Dirac fermions on a $\pi$
flux lattice}

\author{Huaiming Guo$^{1,2}$ }
\author{Ehsan Khatami$^{3}$}
\author{Yao Wang$^{4,5}$}
\author{Thomas P. Devereaux$^{5,6}$}
\author{Rajiv R. P. Singh$^{2}$}
\author{Richard T. Scalettar$^{2}$}

\affiliation{$^1$Department of Physics, Key Laboratory of Micro-Nano Measurement-Manipulation and Physics (Ministry of Education), Beihang University,
Beijing, 100191, China}
\affiliation{$^2$Physics Department, University of California, Davis,
Ca 95616, USA}
\affiliation{$^3$Department of Physics and Astronomy,
San Jose State University, San Jose, CA 95192, USA}
\affiliation{$^4$Department of Applied Physics,
Stanford University, California 94305, USA}
\affiliation{$^5$SLAC National Accelerator Laboratory,
Stanford Institute for Materials and Energy Sciences,
2575 Sand Hill Road, Menlo Park, California 94025, USA}
\affiliation{$^6$Geballe Laboratory for Advanced Materials,
Departments of Physics and Applied Physics, Stanford University,
Stanford, California 94305, USA}

\begin{abstract}
The pairing symmetry of interacting Dirac fermions on the $\pi$-flux
lattice is studied with the determinant quantum Monte Carlo and
numerical linked cluster expansion methods. The $s^*$- (i.e. extended $s$-) and $d$-wave
pairing symmetries, which are distinct in the conventional square
lattice, are degenerate under the Landau gauge.  We demonstrate that the
dominant pairing channel at strong interactions is an unconventional $ds^*$-wave
phase consisting of alternating stripes of $s^*$- and $d$-wave phases.
A complementary mean-field analysis shows that while the $s^*$-
and $d$-wave symmetries individually have nodes in the energy spectrum,
the $ds^*$ channel is fully gapped. The results represent a new
realization of pairing in Dirac systems, connected to the problem of
chiral $d$-wave pairing on the honeycomb lattice, which might be more
readily accessed by cold-atom experiments.
\end{abstract}


\maketitle

\section{Introduction}
One of the dominant themes of condensed matter physics concerns unconventional
superconductivity.  Beginning with the heavy fermions and cuprates, where antiferromagnetic interactions are believed to mediate
$d_{x^2-y^2}$-wave (for simplicity, referred to below as $d$-wave) pairing ~\cite{sc1,sc2}, to $s_{\pm}$ order in the
iron-pnictides~\cite{chubokov08,si16}, growing classes of materials
including, for example, Sr$_2$RuO$_4$, BC$_3$, SrPtAs, MoS$_2$ and
Na$_x$CoO$_2$ have been suggested to host pairing states in which there
are additional broken parity, translation, time-reversal, and rotation
symmetries.

One of the most well-studied of these systems is doped graphene, where
recent theoretical work has demonstrated a chiral $d$-wave
superconducting state~\cite{rev}.  The qualitative explanation for this unconventional phase lies in the fact that the
$d_{x^2-y^2}$ and $d_{xy}$ pairing symmetries belong to the same
irreducible $E_{2g}$ representation of the honeycomb geometry, leading
to the possibility that a complex combination might be energetically favored.
However, determining the correct low temperature superconducting
symmetry, especially in competition with other types of spin density
wave and charge density wave order, and the presence of significant
electron correlation, requires the use of the most discerning analytic
and numeric approaches.  Indeed, methods ranging from mean-field
theory~\cite{meanfield1,meanfield2}, to functional renormalization
group~\cite{nphy,renorm1,renorm2,renorm3} and high-precision numerical
simulations~\cite{vqmc,dqmc,dmrg,ranying,dca} have been applied to the
problem.

The low-energy excitations in graphene are Dirac fermions, which possess
a linear energy dispersion and density of states.  In addition to the
possibility of chiral $d$-wave pairing, these features lead to a variety of
further unusual phenomena ~\cite{graphene}. Given the tremendous interest
in the emergent properties of Dirac fermions, it is natural to examine
their behavior in the absence of graphene's six-fold rotational
symmetry, and with different dispersion relations.

In this manuscript, we employ two unbiased numerical methods, the determinant
quantum Monte Carlo (DQMC)~\cite{blankenbecler81} and the numerical linked-cluster
expansion (NLCE)~\cite{nlce1,nlce4}, to address this important issue by examining the
pairing symmetry of the $\pi$-flux phase square lattice, which, like
graphene, also hosts Dirac fermions.  Originally proposed by Affleck and
Marston to describe the pseudogap regime of the high-$T_c$
cuprates~\cite{affleck}, the $\pi$-flux phase has recently been shown
to be generated spontaneously with dynamical fermions coupled to a
$\mathbb{Z}_2$ gauge theory in $(2+1)$ dimensions~\cite{vishwanath}.  Our
key findings are the following:
(i) Our numerical results paint a consistent picture of the
dominant pairing symmetry, which is found to
be formed by pair creation with alternating stripes of
extended $s$-(denoted as $s^*$-) and $d$-wave symmetries;
(ii) This mixed structure originates in a symmetry linking the
two pairing orders, and possesses a full gap, unlike the
individual pieces;
(iii) Superconductivity is most robust at
intermediate values of the on-site repulsion $U$;
and
(iv) Mean-field theory confirms the basic qualitative picture
coming out of the DQMC/NLCE calculations.
In the conclusions we will also address the possibility of engineering
such lattices using optically trapped atomic systems.

\begin{figure}[t!]
\centering \includegraphics[width=8cm]{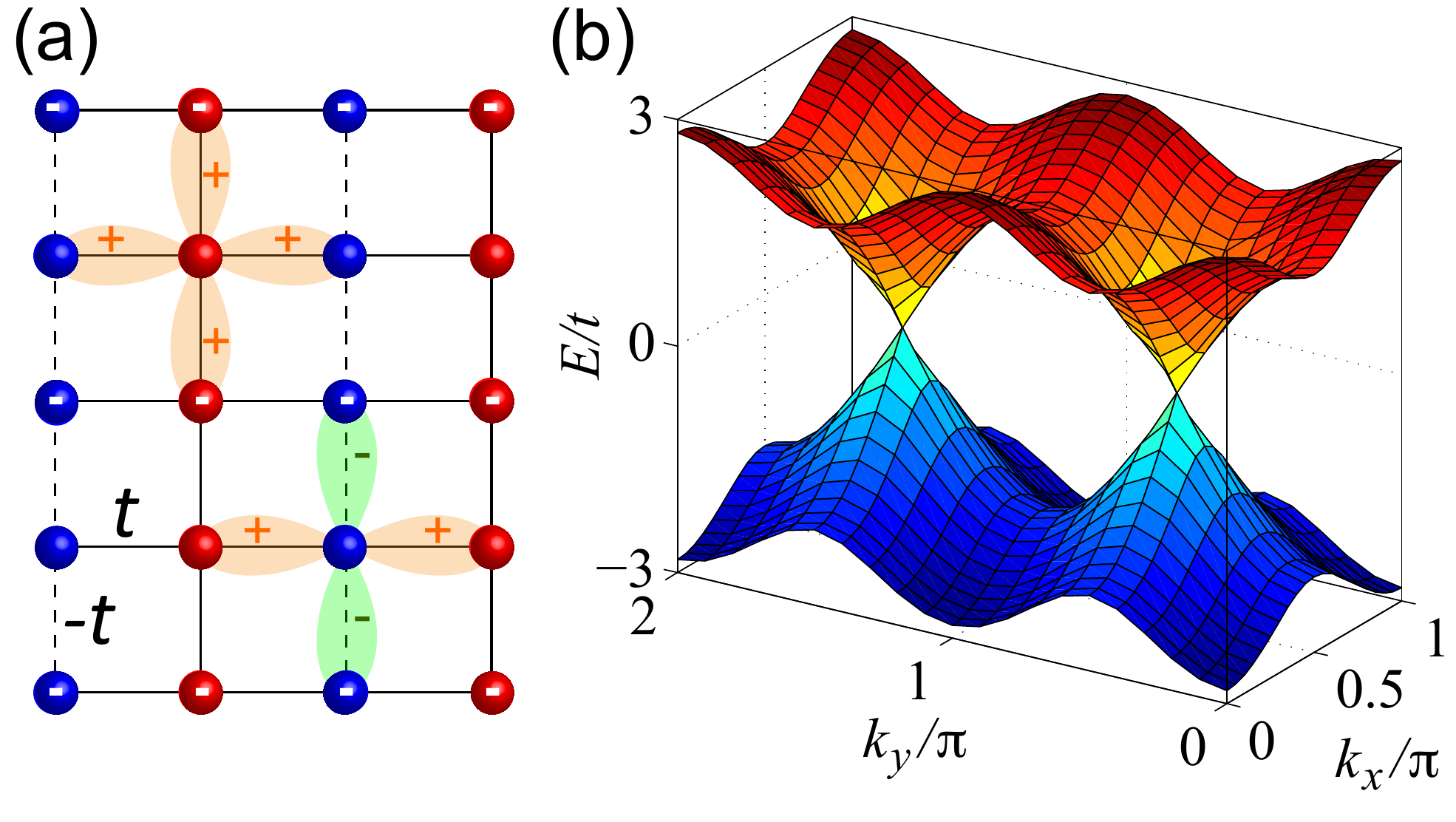} \caption{(a)
The $\pi$-flux lattice in the Landau gauge. The solid (dashed) lines
represent positive (negative) hoppings.  The $ds^*$-wave
pairing symmetry is schematically shown. A gauge transformation
on sites marked by the white bars shows that $s^*$- and $d$-waves
are equivalent. (b) The noninteracting energy spectrum, which shows that the
system is a semi-metal with two inequivalent Dirac points.
The corresponding density of state is linear for low energies and has
a Van Hove singularity at $E/t=2$.}
\label{fig1}
\end{figure}

\section{Model and method}
We consider a Hubbard Hamiltonian describing
interacting Dirac fermions in a $\pi$-flux model on a square lattice
where each plaquette is threaded with half a flux quantum,
~\cite{piflux1,piflux2}
$\frac{1}{2}\Phi_{0} =hc/(2e)$,
\begin{equation}\label{eq1}
H=\sum_{\langle lj \rangle \sigma}
t_{lj}e^{i\chi_{lj}}c^\dag_{j\sigma}c^{\phantom{\dag}}_{l\sigma}+U\sum_{i}(n_{i\uparrow}-\frac{1}{2})(n_{i\downarrow}-\frac{1}{2}),
\end{equation}
where $c^\dag_{j\sigma}$ and
$c^{\phantom{\dag}}_{j\sigma}$
are the creation and annihilation operators, respectively, at site $j$
with spin $\sigma=\uparrow, \downarrow$.
The hopping amplitudes between the nearest-neighbor sites
$l$ and $j$ are $t_{lj}=t$, which we set to 1 as the unit of energy throughout
our paper, and $\chi_{lj}$ is the Peierls phase arising from
the magnetic flux
$\chi_{lj}=\frac{2\pi}{\Phi_{0}}\int_{{\bf x}_l}^{{\bf x}_j}
{\bf A}\cdot d{\bf x}$ with ${\bf A}$ the vector potential.
In the Landau gauge we have ${\bf A} = \frac{1}{2} \Phi_{0}(0, x)$
and the Peierls phase is given by $\chi_{j,j+\hat{x}}=0,  \,\,
\chi_{j,j+\hat{y}}=\pi j_x$. The resulting hopping pattern is shown
in Fig.~\ref{fig1}(a).
The specific form of
$\chi_{lj}$ is gauge-dependent, allowing for different
choices of the Peierls factors~\cite{sup}. In the following, results
are based on the geometry of Fig.~\ref{fig1}(a). We have verified
that results for other gauge choices are consistent.

The lattice in Fig.~\ref{fig1}(a) has a two-site unit cell.  In
reciprocal space, with the reduced Brillouin zone
$(|k_x| \leq \pi/2, |k_y| \leq \pi)$,
the Hamiltonian can be written as $H_0=\sum_{\bf{k}\sigma}
\psi_{\bf{k}\sigma}^{\dagger} {\cal H}_0(\bf{k})
\psi_{\bf{k}\sigma}^{\phantom{\dagger}}$
with $\psi_{\bf{k}\sigma}^{\phantom{\dagger}}=(c_{{\bf k}\sigma}^{1},c_{{\bf k}\sigma}^{2})^{T}$
and ${\cal H}_0({\bf k})=2t\cos{k_x}\sigma_{x}-2t\cos{k_y}\sigma_{z}$,
with $\sigma_{x,z}$ the Pauli matrices. The energy spectrum is given by
$E_{\bf k} = \pm \sqrt{4t^{2}(\cos^2 k_x+\cos^2 k_y)}$.
The noninteracting system is a semi-metal with two inequivalent
Dirac points at ${\bf K}_{1,2}=(\pi/2,\pm \pi/2)$ as shown in Fig.~\ref{fig1}(b).

The interacting $\pi$-flux model is solved numerically by
means of the DQMC and the NLCE methods. We also validate our results
using exact diagonalization (ED) for a $4\times 4$ lattice~\cite{sup}. In DQMC, one decouples the
on-site interaction term through the introduction of an auxiliary
Hubbard-Stratonovich field, which is integrated out stochastically.
The only errors are those associated with the statistical sampling,
finite spatial lattice size, and the inverse temperature discretization.
All are well-controlled in the sense that they can be systematically
reduced as needed, and further eliminated by appropriate extrapolations.
At half-filling (average density of one fermion per site), we have access to low-temperature
results, necessary to determine the pairing symmetry. Away from
half-filling and in the presence of the ``sign problem"~\cite{loh90,vlad2015} in the DQMC, we
can access temperatures down to $T \sim 0.4$.
The DQMC simulations are carried out on a $12 \times 12$ system,
which is large enough to have negligible finite-size
effects for the temperatures studied here~\cite{sup}.
Results represent averages of $10$ independent runs
with $10000$ sweeps each.

In the NLCE, properties in the thermodynamic limit are expressed in terms of
contributions from small clusters that can be embedded in the lattice. The latter are obtained via ED.
We use a NLCE for the square lattice,
modified to fit in the reduced symmetry of the $\pi$-flux
model, and carry out the expansion up to the 8th order~\cite{nlce4,checkerboard}.
NLCE is error free in the temperature region of convergence and can be used
to gauge systematic errors in  DQMC in the common region of validity.
Here we show both the bare results and those obtained after Euler resummation~\cite{sup}.

The quantity on which we focus~\cite{sup}
is the pairing structure factor, $S^{\alpha}({\bf q})=\sum_{\bf r}e^{i{\bf q}\cdot {\bf r}}P^{\alpha}({\bf r})$,
where
$P^{\alpha}({\bf r}_{ij})=\langle \Delta_{i}^{\alpha \dagger}(0)
\Delta_{j}^{\alpha}(0)+\Delta_{i}^{\alpha}(0)
\Delta_{j}^{\alpha \dagger}(0)\rangle$
is the equal-time pair-pair correlation function. The general
(time dependent) pairing operator
is defined as $\Delta_{i}^{\alpha}(\tau)=
\sum_{j}f_{ij}^{\alpha}\,e^{\tau H}c_{i\uparrow}c_{j\downarrow}
e^{-\tau H}$ with $f_{ij}^{\alpha}=\pm 1$ for the bond
connecting $i$ and $j$, depending on the pairing symmetry
$\alpha$.
The $\Delta_{ds^*}$ operator which proves to be dominant
on the $\pi$-flux phase lattice possesses $d$-wave
phases ($f_{ij}=+1$ for $j=i \pm \hat x$ and
$f_{ij}=-1$ for $j=i \pm \hat y$)
for sites on vertical stripes of the lattice
with $i_x$ odd, and $s^*$-wave symmetry
($f_{ij}=+1$ for both $j=i \pm \hat x$ and $j= i \pm \hat y$)
for $i_x$ even.
As we shall show below, this symmetry has a larger
superconducting response than more conventional singlet pairings in the
 $s^*$, $d_{x^2-y^2}$, $s_{xy}$, and $d_{xy}$ channels, and triplet pairings in
 $p_x$, $p_y$, and $p_{xy}$ channels\cite{note1}.

Here we consider only the uniform pairing structure factor, $S^{\alpha}({\bf q}=0)$ and its correlated part
, $S_{corr}^{\alpha}$, obtained by subtracting off the uncorrelated parts from $S^{\alpha}$.
One can also analyze the uniform pairing {\it susceptibility},
\begin{eqnarray}\label{eq4}
\chi^{\alpha}({\bf q}=0)=\frac{1}{N}\int_{0}^{\beta}d\tau \sum_{ij}\langle \Delta^{\alpha}_{i}(\tau)\Delta^{\alpha\dagger}_{j}(0)\rangle
\,,
\end{eqnarray}
which probes the decay of pairing correlations in the
imaginary time as well as spatial directions.
As with the structure factor,
a subtraction of the uncorrelated pieces of $\chi^\alpha$
can be used to evaluate the pairing vertex~\cite{sc2}.
Susceptibilities generally have stronger signals in ordered phases~\cite{sc3}.
However they also have larger error bars in the DQMC and are
substantially more costly to compute.

\section{Superconducting pairing symmetry}
Spin fluctuations play an important role in pairing in Hamiltonians with
repulsive electronic interactions, both competing with superconductivity
at half-filling and providing the `pairing glue' upon doping.
Unlike in the square lattice model with equal hoppings,
for which the critical interaction $U_c=0$,
antiferromagnetic (AF) order in the $\pi$-flux lattice with Dirac fermions
only develops above $U_c = 5.64\pm 0.05$~\cite{pmag1,pmag2,pmag3,pmag4}.
However, we find that short-range AF correlations behave very similarly in the
two models, suggesting that magnetic pairing mechanisms might be equally
robust in the two cases~\cite{sup}.

\begin{figure}[htbp]
\centering \includegraphics[width=9cm]{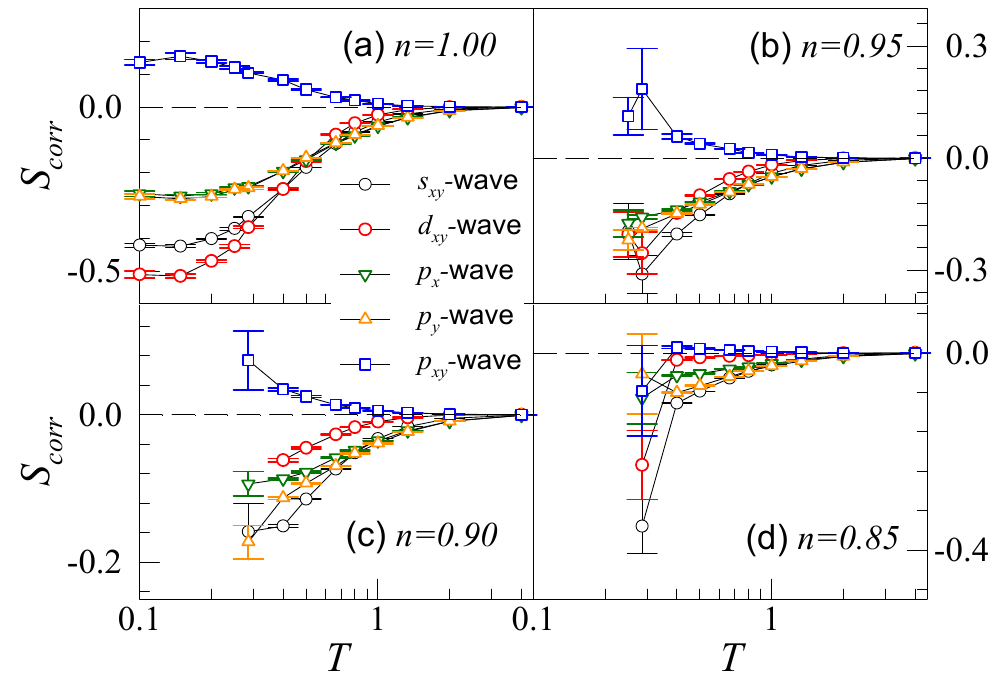}
\caption{DQMC results
for the ${\bf q}=0$ (uniform) $s_{xy}$-wave, $d_{xy}$-wave,
$p_x$-wave, $p_y$-wave and $p_{xy}$-wave pairing structure factors
as a function of temperature.
Here $U=8t$ and the densities are: (a) $n=1.00$;
(b) $n=0.95$; (c) $n=0.90$; (d) $n=0.85$.
All channels are repulsive except for weakly attractive $p_{xy}$.}
\label{fig2}
\end{figure}

\begin{figure}[htbp]
\centering \includegraphics[width=9cm]{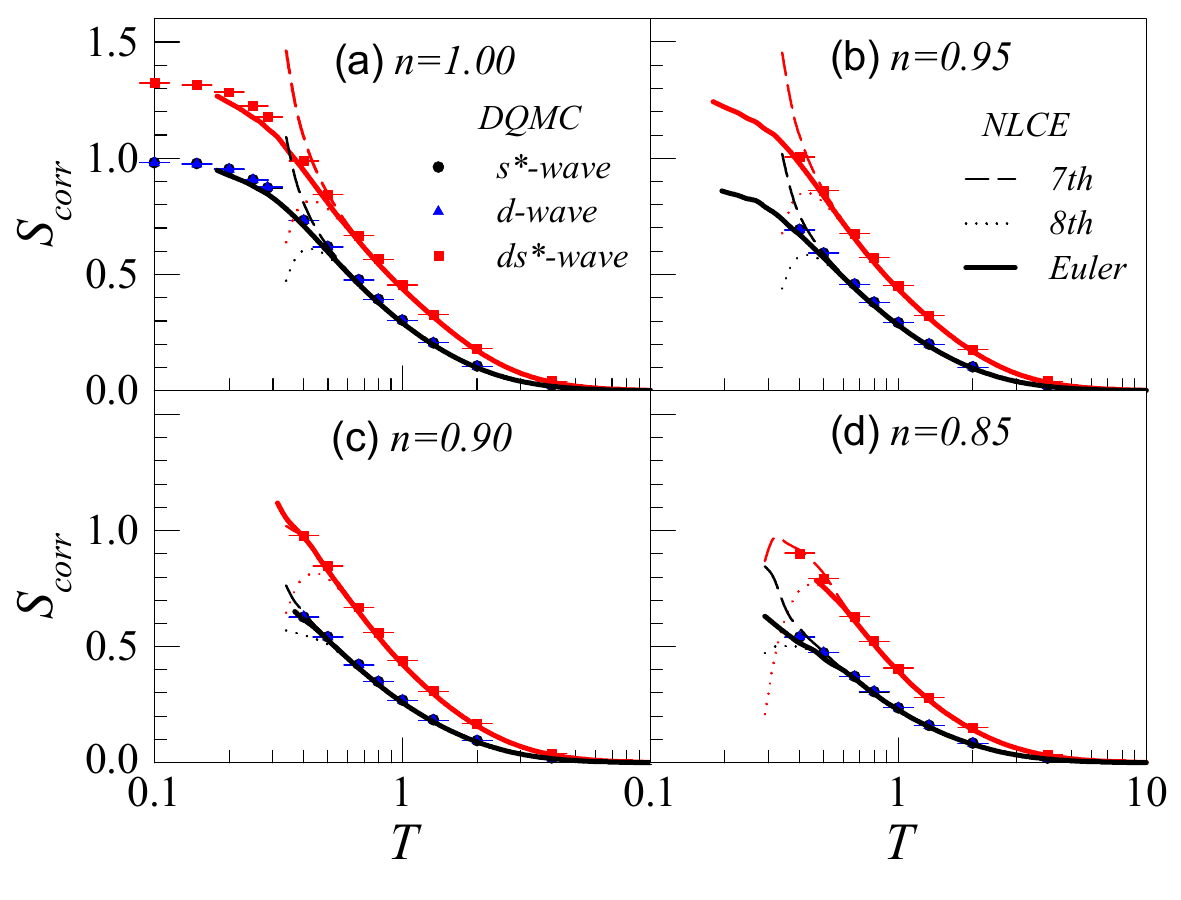}
\caption{The $ds^*$-wave, uniform $d$-wave and $s^*$-wave pairing
structure factors vs
temperature for $U=8t$ at densities $n=1.00,0.95,0.90,0.85$. $s^*$-wave and $d$-wave are identical to the accuracy of
our calculations. Symbols are from the DQMC. Thin dashed and dotted lines are bare NLCE results for the 7th and 8th orders, respectively. Thick solid lines are results after the Euler resummation~\cite{sup}.}
\label{fig3}
\end{figure}

In Fig.~\ref{fig2}, we show the correlated part of the uniform structure factor
for several of the pairing symmetries, at various dopings for $U=8$.
DQMC can access low temperatures at half-filling, but is blocked
by the `sign problem' in doped systems~\cite{loh90}.  Nevertheless, the
increasingly negative correlated structure factors in the $p_x, p_y,
s_{xy}, d_{xy}$ modes offer compelling evidence that these symmetries
are suppressed.  For the $s_{xy}$ and $d_{xy}$ this can be understood as
a consequence of the tendency towards AF order, with parallel spin
fermions on next-nearest-neighbor (NNN) sites at odds with the presence
of a singlet pair.  The $p_{xy}$ mode is attractive, but its value is
much smaller than $s^*$ and $d$-wave pairing (Fig.~\ref{fig3}).

We find that $s^*$-, $d$-, and $ds^*$-wave pairings are
an order of magnitude larger than $p_{xy}$-wave, and that $ds^*$-wave pairing
is dominant in all parameter regions. By symmetry,
$s^*$-, $d$-wave channels are equivalent in this model. This can be seen
as follows:
The $\pi$-flux lattice under Landau gauge belongs to the group $D_{2h}$.
Among the irreducible representations for the group with $k_z=0$,
$A_{1g}$ has the basis function $k_x^2$ or $k_y^2$, which are
independent.  The $s^*$ ($d$)-wave is a linear combination of the two
basis functions $k_x^2+k_y^2$ ($k_x^2-k_y^2$); thus they are not
necessarily equal from the point of view of the crystal symmetry group. However
gauge symmetry, a hidden symmetry underlying the Hamiltonian, enforces
their equivalence.  This can be directly seen by performing a
transformation on the sites marked by white bars in
Fig.~\ref{fig1}(a), $c_{i,\sigma} (c^{\dagger}_{i,\sigma})\rightarrow
-c_{i,\sigma} (-c^{\dagger}_{i,\sigma})$, under which the Hamiltonian
remains unchanged while the uniform $s^*$-wave pairing becomes $d$-wave
(or vice versa). This equivalence is confirmed within machine precision in the
NLCE.

\begin{figure}[htbp]
\centering \includegraphics[width=8cm]{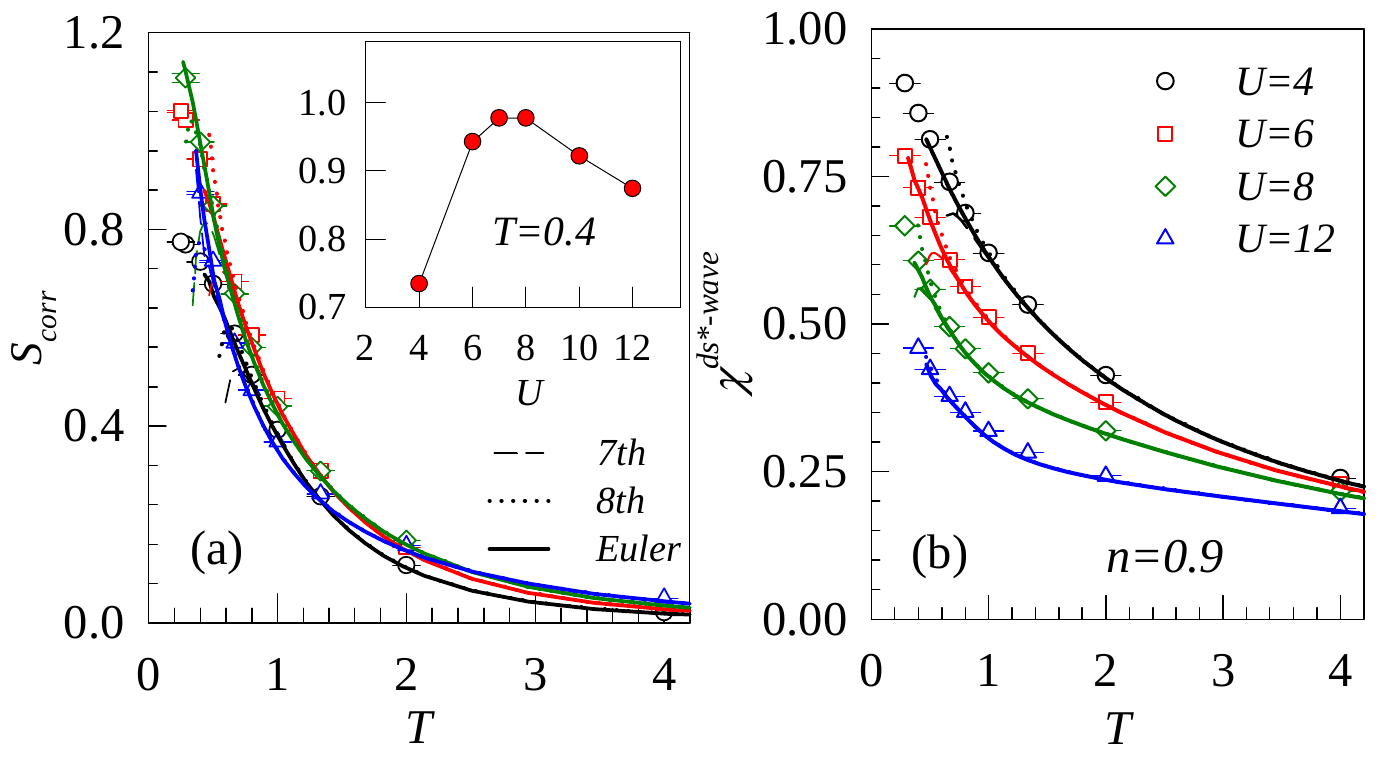}
\caption{(a):
Temperature dependence of $ds^*$-wave pairing structure factor
at density $n=0.9$ for different values of the interaction. The inset
shows the structure factor vs $U$ at a fixed temperature $T=0.4$.
A maximum is present at intemediate coupling.
Symbols and lines in the main panels are the same as in Fig.~3.
(b): The $ds^*$-wave pairing susceptibility as a function of the
temperature at $n=0.9$ for different values of $U$.
}
\label{fig4}
\end{figure}

As shown in Fig.~\ref{fig3}, the $ds^*$-wave pairing has the largest
correlated structure factor for a range of dopings about half-filling.
Results from NLCE and DQMC are in very good agreement and point to a
saturation of $S_{corr}$ at low temperatures at zero and 5\% doping ($n=0.95$).
However, we are limited to relatively high temperatures at the other two doping values
shown in Fig.~\ref{fig3}, where $S_{corr}$ continues to increase as $T$ is lowered.
We focus on $n=0.90$, and plot $S_{corr}$ vs temperature for $U=4,6,8$ and $12$
in Fig.~\ref{fig4}(a). At low temperature, the structure
factor quickly rises as $U$ increases from $U=4$, reaches a maximum
in the intermediate-coupling region, and then slowly decreases. Figure \ref{fig4}(b) shows the susceptibility
$\chi$ vs temperature for different interaction strengths at $n=0.90$.
For large $U$, there is a trend for the susceptibility to rapidly
increase at low temperatures.
The full $ds^*$-wave susceptibility shows a clear enhancement
over its uncorrelated value, implying the pairing interaction is attractive.
As in Fig.~\ref{fig3}, the results from NLCE match well with DQMC
in Fig.~\ref{fig4}, indicating that systematic errors are not significant at the
accessible temperatures.

Magnetic orders may compete with the superconductivity discussed above. We can not rule out the possibility of a magnetic ground state, however, lack of nesting, resulting in $U_c>0$ for LRAFO, and the incommensurate filling make the magnetic order less competitive.

\begin{table}
\begin{tabular}{|c|c|c|}
  \hline
  $s^{*}$ & $\lambda=\cos k_y \pm |\cos k_x|$ & $P_s^*({\bf k})=2\Delta \cos k_y \tau_x\otimes I$\\
  \hline
  $d$ & $\lambda=-\cos k_y \pm |\cos k_x|$ & $P_d({\bf k})=-2\Delta \cos k_y \tau_x\otimes I$\\
  \hline
  $ds^*$ & $\lambda^2=\cos k_x^2+\cos k_y^2$ & $P_{ds^*}({\bf k})=2\Delta \cos k_y \tau_x\otimes \sigma_z$\\
  \hline
 \end{tabular}
\caption{The character value $\lambda$ of the gap matrix and $P_{\alpha}$ in Eq.(7) for three typical pairings.} \label{z1table}
\end{table}

\section{Mean-field description of the superconducting state}
\begin{figure}[htbp]
\centering \includegraphics[width=9cm]{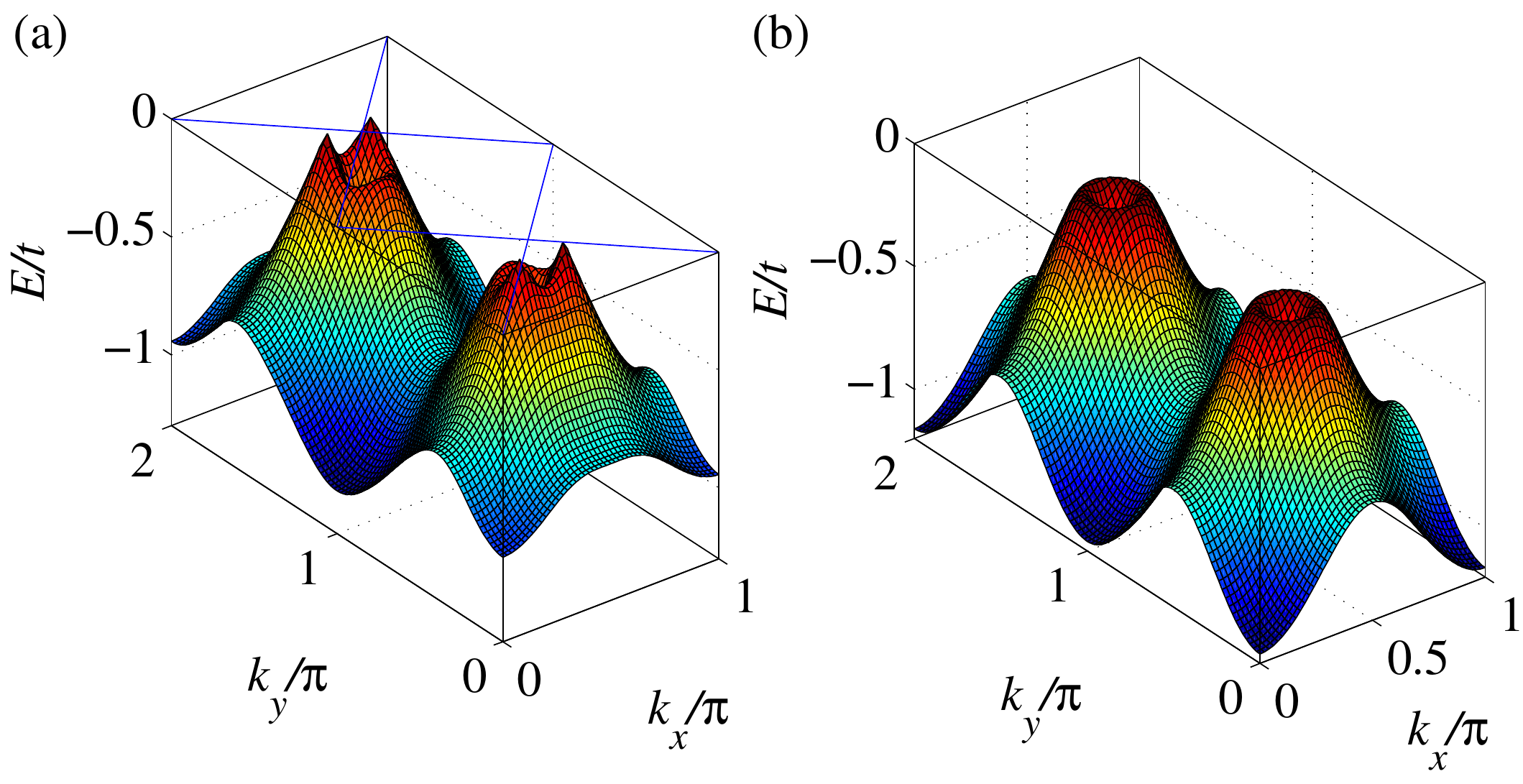} \caption{The lower energy
dispersion within the mean-field theory
near the Fermi energy for $s^*$-wave (or $d$-wave)
(a) and $ds^*$-wave (b). Here the parameters are $\mu=0.8, \Delta=0.2$.}
\label{fig5}
\end{figure}

To study the physical properties of the possible superconducting states
further, we analyze the gap function, $\Delta^{\alpha}=\sum_{i} \Delta_{i}^{\alpha}(0)=
\sum_{\bf k} \Phi^{T}_{\uparrow}({\bf k})D^{\alpha}\Phi_{\downarrow}(-{\bf k})$, where
\begin{equation}
D^{\alpha}=\left(
          \begin{array}{cc}
            \gamma \cos k_y & \cos k_x \\
            \cos k_x & \beta \cos k_y \\
          \end{array}
        \right),
\end{equation}
and $\Phi_{\sigma}({\bf k})=(c_{A,{\bf k}\sigma},  c_{B,{\bf
k}\sigma})$ and $\gamma,\beta=1(-1)$ for $s^{*}(d)$-wave pairing on each
site.  The character values $\lambda$ of the gap matrix are shown in Table
I. $s^*$- and $d$-wave have nodes along the blue lines in Fig.~\ref{fig5},
while $ds^*$-wave is fully gapped.

A mean-field analysis of the superconducting spectrum provides a qualitative check on the DQMC and NLCE results reported above.
The nonlocal pairing channels can not be decoupled from the on-site
Hubbard term. However at large $U$, the low-energy physics can be
captured within the $t$-$J$ model~\cite{tjmodel}. The single-occupancy restriction is dealt with in an average way by the use of statistical weighting factors $t_{eff}=\frac{2\delta}{1+\delta}t$ and $J_{eff}=\frac{4}{(1+\delta)^2}J$ with $\delta$ the doping level and the coupling constant $J = \frac{4t^2}{U}$. The
Heisenberg coupling is expressed in terms of the spin-singlet operator, $J_{eff}({\bf
S}_{i}\cdot{\bf S}_{j}-\frac{1}{4}n_{i}n_{j})=-J_{eff} h^{\dagger}_{ij}h_{ij}$
with $h^{\dagger}_{ij}=\frac{1}{\sqrt{2}}(c^{\dagger}_{i\uparrow}
c^{\dagger}_{j\downarrow}-c^{\dagger}_{i\downarrow}c^{\dagger}_{j\uparrow})$,
with $i$ and $j$ near neighbors.
The mean-field parameter is $\Delta_{ij}=-J_{eff}\langle
\, h_{ij}\, \rangle/\sqrt{2}$.
In the basis $\psi_{\bf k}=(c_{1,{\bf k}\uparrow},c_{2,{\bf
k}\uparrow},c_{1,{\bf -k}\downarrow}^{\dagger},c_{2,{\bf
-k}\downarrow}^{\dagger})^{T}$, we arrive at the mean-field Hamiltonian: $H_{MF}=\sum_{\bf k} \psi^{\dagger}_{\bf k}{\cal H}_{MF}({\bf
k})\psi_{\bf k}+E_0$ with ${\cal H}_{MF}({\bf k})=t\cos k_x \tau_z\otimes \sigma_x-t\cos k_y \tau_z \otimes \sigma_z
-\frac{\mu}{2} \tau_z\otimes I +2\Delta \cos k_x \tau_x\otimes \sigma_x+P_{\alpha}({\bf k})$ and a constant term $E_0=4N\frac{\Delta^2}{J_{eff}}$.
 The ground state is then obtained by
minimizing the free energy with respect to the order parameter $\Delta$ and doping $\delta$, which yields two self-consistent equations. After a numerical self-consistent
iteration, we find that the order parameter $\Delta$ of the $ds^*$-wave pairing has larger values for the low doping levels, implying it is dominating in the ground state.

It is also straightforward to obtain the energy dispersion. We plot the bands
near the Fermi energy in Fig.~\ref{fig5}. The $s^*$- or $d$-wave pairing
states are seen to have nodes, while the $ds^*$-wave state is fully gapped.
A qualitative argument for the dominance of $ds^*$ pairing is the
following:  As emphasized by Scalapino~\cite{sc2}, the presence of
a self-consistent solution of the
gap equation
$\Delta_k=-\sum_{k'} \Gamma_{kk'} \big(\Delta_{k'}/2E_{k'} \big)
\, {\rm tanh}(E_{k'}/2T)$, where $E_{k}$ is the superconducting quasiparticle dispersion,
for repulsive interactions $\Gamma_{kk'}$ necessitates a change in
sign of $\Delta_k$, and hence the presence of nodes.
However nodes reduce the overall energy lowering due to gap formation in
the superconducting states.  As a consequence, a symmetry which enables
a non-trivial self-consistent solution, while leaving the gap everywhere
large, is energetically preferred.

\section{Conclusions}
Pairing in the Hubbard model on a $\pi$-flux lattice was studied using
exact/large-scale numerical methods. The $s^*$- and $d$-wave symmetries,
which are distinct in the most commonly studied square lattice, are equivalent
under the Landau gauge.  Both DQMC and NLCE indicate that the dominating
pairing channel at strong interactions is an unconventional $ds^*$-wave, for
which the relative signs of the pairing amplitudes alternate between
$d$-wave and $s^*$-wave patterns on adjacent stripes of the lattice.
Within a mean-field analysis, the $s^*$- or $d$-wave channels can be
shown individually to have nodes while the $ds^*$ channel is fully
gapped.  The results represent a profound extension of studies of
interacting Dirac fermions in graphene by eliminating the specific
symmetries of the honeycomb lattice.  The DQMC studies reported here
cannot access the Van Hove singularity at quarter-filling ($n=0.5$),
where the instability to a chiral $d$-wave state is especially prominent
in graphene~\cite{rev}.  However ED simulations on small lattices show a
sign that the gapless $s^*$- or $d$- channel may dominate there, which
warrants further studies.

Finally, we discuss how this phase might be accessed by
state-of-art cold-atom experiments~\cite{cold1,cold2}.
It is by now well-established that Raman-assisted tunneling, and other
methods, can be used to create effective magnetic fields on optical
lattices\cite{jaksch03,gerbier10,mueller04,aidelsburger11,aidelsburger13,miyake13,cold1,cold2},
as well as more complex (non-Abelian) artificial gauge
fields\cite{spielman16}.  The hybridization pattern of Fig.~\ref{fig1}
corresponds to alternating $\pm\pi$ magnetic flux on adjacent
vertical stripes of the lattice, in precisely the geometry of
Ref.~[\onlinecite{aidelsburger11}], which achieved $\phi=\pm \pi/2$
flux, similarly alternating along the $\hat x$ direction.  As discussed
there, changing the wavelength of the Raman lasers, or the angle between
them, allows for generally tunable $\phi$. The
pattern proposed here has already been realized for bosons~\cite{metai17}. Recent advances in high-resolution control of the confining potential, resulting in flat regions~\cite{mazurenko17}, can mitigate issues related to density inhomogeneity. These could, then, provide a precise and
well-controlled realization of the unconventional $ds^*$ pairing symmetry
described here.

\section{Acknowledgements}
The authors thank C.~C.~Chang, Z. X. Li, W.~Pickett, S.~Raghu and F. Zhang for helpful
discussions.  H.G.~acknowledges support from China Scholarship Council and NSFC under Grant No. 11774019. E.K. is supported
by NSF under Grant No. DMR-1609560.
Y.W. and T.P.D. are supported by DOE grant No. DE-AC02-76SF00515. A portion of the computational work was performed using the resources of the National Energy Research Scientific Computing Center supported by DOE grant No. DE-AC02-05CH11231.
R.R.P.S. is supported by NSF under Grant No. DMR-1306048.
The work of R.T.S.~is supported by DOE grant No. DE-SC0014671.

\renewcommand{\thefigure}{A\arabic{figure}}
\setcounter{figure}{0}
\appendix
\section{Context of Pairing Symmetry}
\begin{figure}[htbp]
\centering \includegraphics[width=7cm]{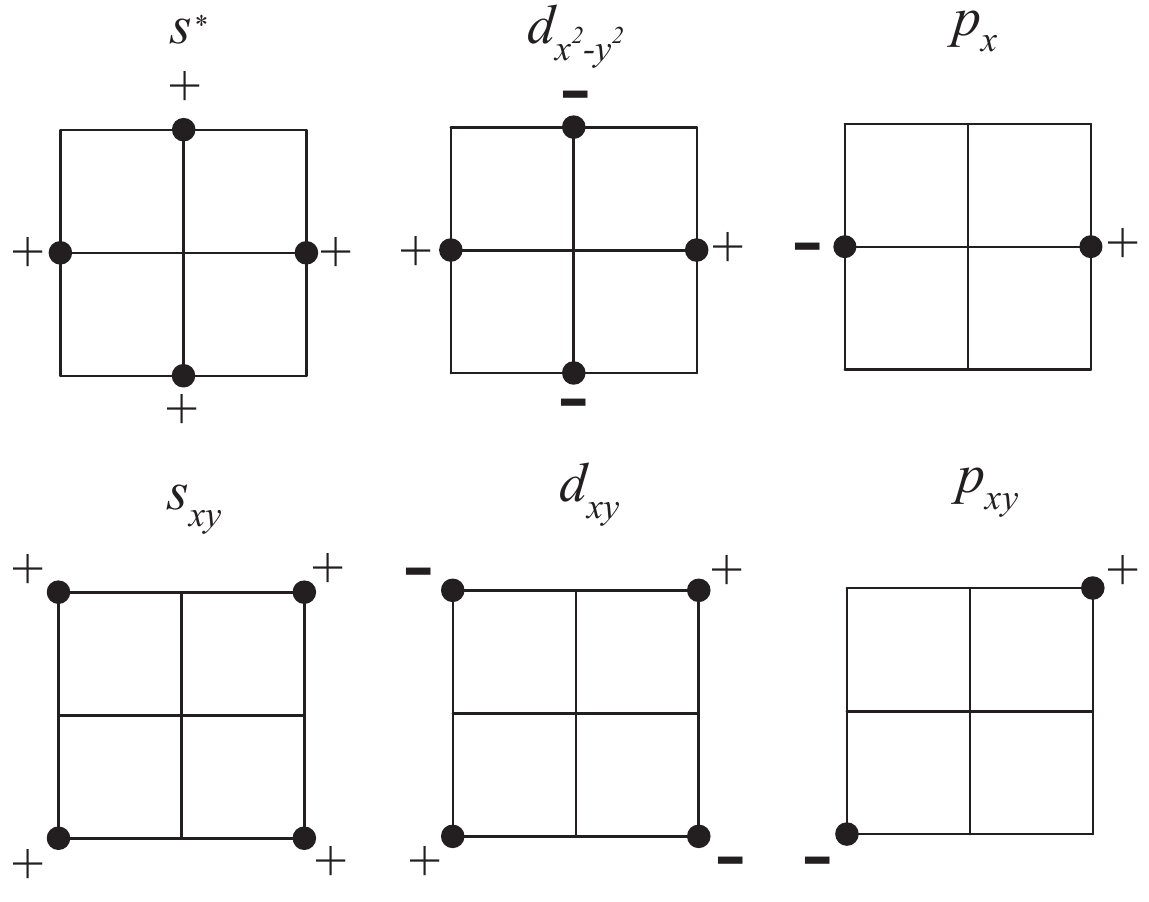} \caption{
Six of the nine the pairing symmetries
available when the down spin fermion is created on a 3x3 lattice
about the location of the up spin fermion at the center.
On-site $s$-wave, where the down spin fermion
is created at the same point as the up spin fermion,
is not shown, nor are $p_y$ and $p_{yx}$, which are just
$90^\circ$ rotations
of the $p_x$ and $p_{xy}$ symmetries
illustrated in the two right-hand panels.
}
\label{figS1}
\end{figure}

In early studies of the Hubbard Hamiltonian on a
square lattice with uniform hopping (no flux), the amplitudes of the
pairing responses of different symmetries were compared \cite{sc1}.
Figure \ref{figS1} shows the real space arrangements of the
wave function of the down spin fermion around the up spin fermion.
These correspond to momentum space pair creation operators,
\begin{align}
\Delta_{\bf k}^{\alpha\dagger} = \sum_{{\bf k}}
\, f_{\bf k}(\alpha) \,
c_{{\bf k}\uparrow}^\dagger
c_{{\bf -k}\downarrow}^\dagger \,\,,
\end{align}
where $\alpha$ distinguishes the different symmetries,
\begin{align}
f_{\bf k}(s)&=1
\hskip1.03in
f_{\bf k}(s^*) = {\rm cos}\,k_x +{\rm cos}\,k_y
\nonumber \\
f_{\bf k}(p_x) & = {\rm sin}\,k_x
\hskip0.50in
f_{\bf k}(d_{x^2-y^2}) = {\rm cos}\,k_x -{\rm cos}\,k_y
\nonumber \\
f_{\bf k}(p_y) &= {\rm sin}\,k_y
\hskip0.70in
f_{\bf k}(d_{xy})  = {\rm sin}\,k_x {\rm sin}\,k_y
\nonumber \\
f_{\bf k}(s_{xy}) &= {\rm cos}\,k_x {\rm cos}\,k_y
\hskip0.34in
f_{\bf k}(p_{xy})  = {\rm sin}\,(k_x + k_y)
\nonumber \\
f_{\bf k}(p_{yx})  &= {\rm sin}\,(k_x - k_y) \,\,.
\end{align}

The $\pi$-flux lattice we consider here, which breaks translational
symmetry in the $\hat x$ direction, allows for more complex symmetries,
including the $ds^*$ arrangement of Fig.~1 of the main text.
As illustrated there, the $ds^*$ symmetry alternates
the $d_{x^2-y^2}$ and $s^*$ patterns of Fig.~S1
as one moves between the $\pm \pi$ flux plaquettes.

\section{Gauge symmetry}
\begin{figure}[htbp]
\centering \includegraphics[width=7cm]{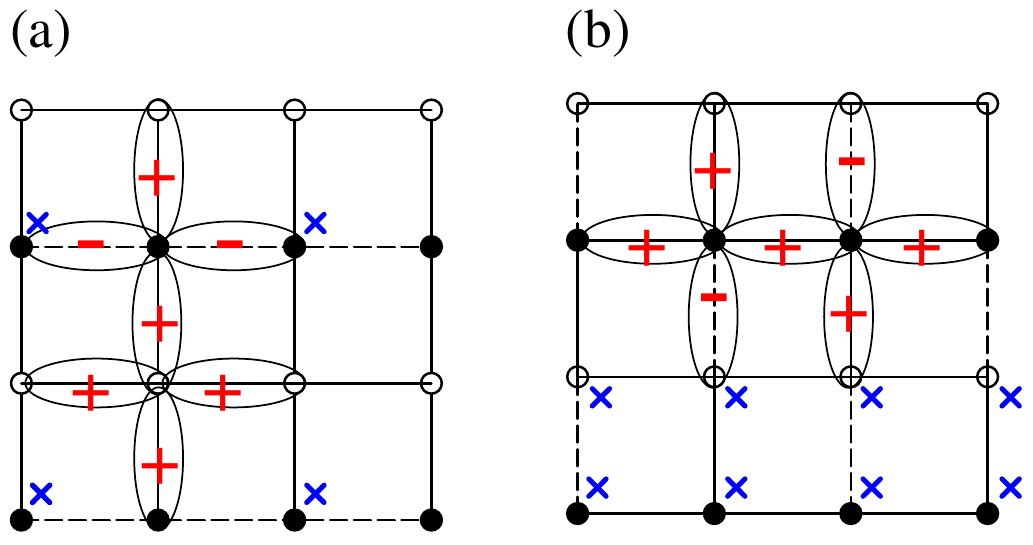} \caption{The $\pi-$flux
lattice under other gauges. The corresponding $ds^*-$wave pairing
symmetry is schemetically shown. The lattice and the pairing symmetry is
transformed from the one under Landau gauge [see Fig.1(a) in the main
text] by a gauge transformation $c_{i,\sigma}
(c^{\dagger}_{i,\sigma})\rightarrow -c_{i,\sigma}
(-c^{\dagger}_{i,\sigma})$ on the sites marked by blue crosses.}
\label{figS2}
\end{figure}

The $\pi$-flux lattice can be realized with different choices
of the hopping, i.e.~with different gauges, as shown in
Fig.~\ref{figS2}.  The hopping pattern is gauge dependent, but so
are the phases of the $ds^*$ hopping.
Two of the alternate choices are shown in Fig.~\ref{figS2}.
In Fig.~\ref{figS2}(a), the vector potential ${\bf A} = -\frac{1}{2}
\Phi_{0}(y, 0)$ is chosen.
As a check on our algorithm, we performed simulations
of these transformed systems, and verified that all
results are consistent with those in the main text.

\begin{figure}[htbp]
\centering \includegraphics[width=8cm]{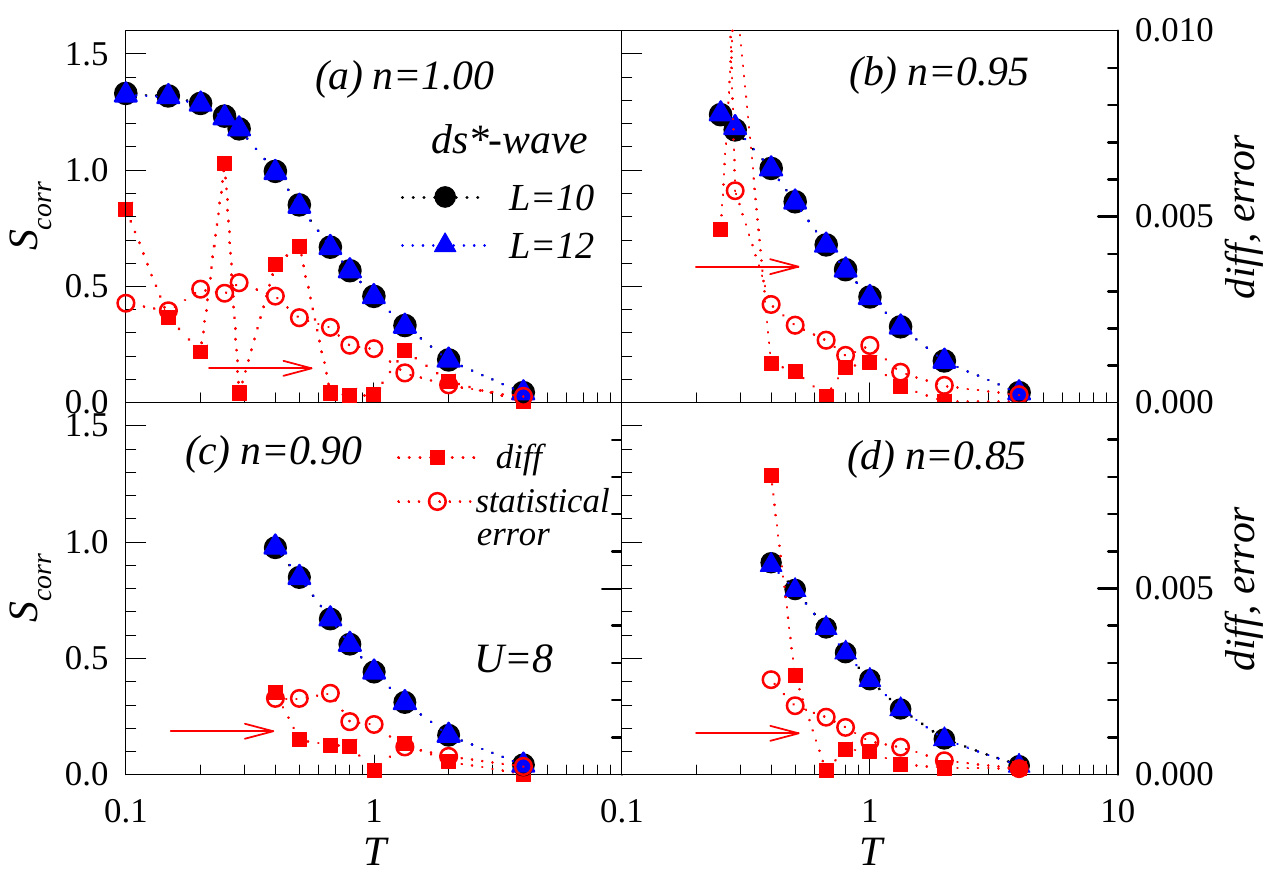}
\caption{The correlated pairing structure factors for two different
lattice sizes, $L=10$ (black circles) and $L=12$ (blue triangles). The absolute difference for the
densities $n=1.00,0.95,0.90,0.85$ at $U=8$ is of order $10^{-3}$, which
is comparable to the statistical error bars (the corresponding axis is marked by the red arrow).}
\label{figS3}
\end{figure}

\section{Finite Size Effects}
In the main text, all DQMC results were obtained on a $12\times 12$
lattice.  In Fig.~\ref{figS3}, we show some results on $10\times 10$
lattice to assess finite size effects. The absolute values of the
differences between the two sizes are of order $10^{-3}$.  We conclude
finite-size effects at the temperatures considered here are
small. This fact is also implied by the agreement between the NLCE
calculations shown in the main text, which represent the
thermodynamic limit, yet match the DQMC results well.

We also note that on the $10\times 10$ lattice, the Dirac points, which
are located at ($\pi/2, \pm \pi/2$), are not captured by the discrete
momenta. As a consequence, the non-interacting band structure is not
degenerate as is the case on $12\times 12$ lattice.  (In one dimension,
at $U=0$, the ground state energy at half-filling of lattices of size
$4n$ and $4n+2$ approach the thermodynamic limit from opposite
directions owing to the presence/absence of $k$ points at the Fermi
surface).  Thus the agreement between the 10x10 and 12x12 lattices is an
even more strict validation that finite size effects are under good
control.  In general, for Hubbard Hamiltonians without any threading
flux, a good rule of thumb\cite{mag4} is that
`shell effects' associated with the discrete momentum grid tend to be
noticable only for $U/t \lesssim 2$ on lattices of the sites studied here.  Above this value, the interaction
sufficiently smears the finite momentum grid to eliminate size effects.

\section{Exact Diagonalization Benchmarks}
\begin{figure}[h!]
\centering \includegraphics[width=8cm]{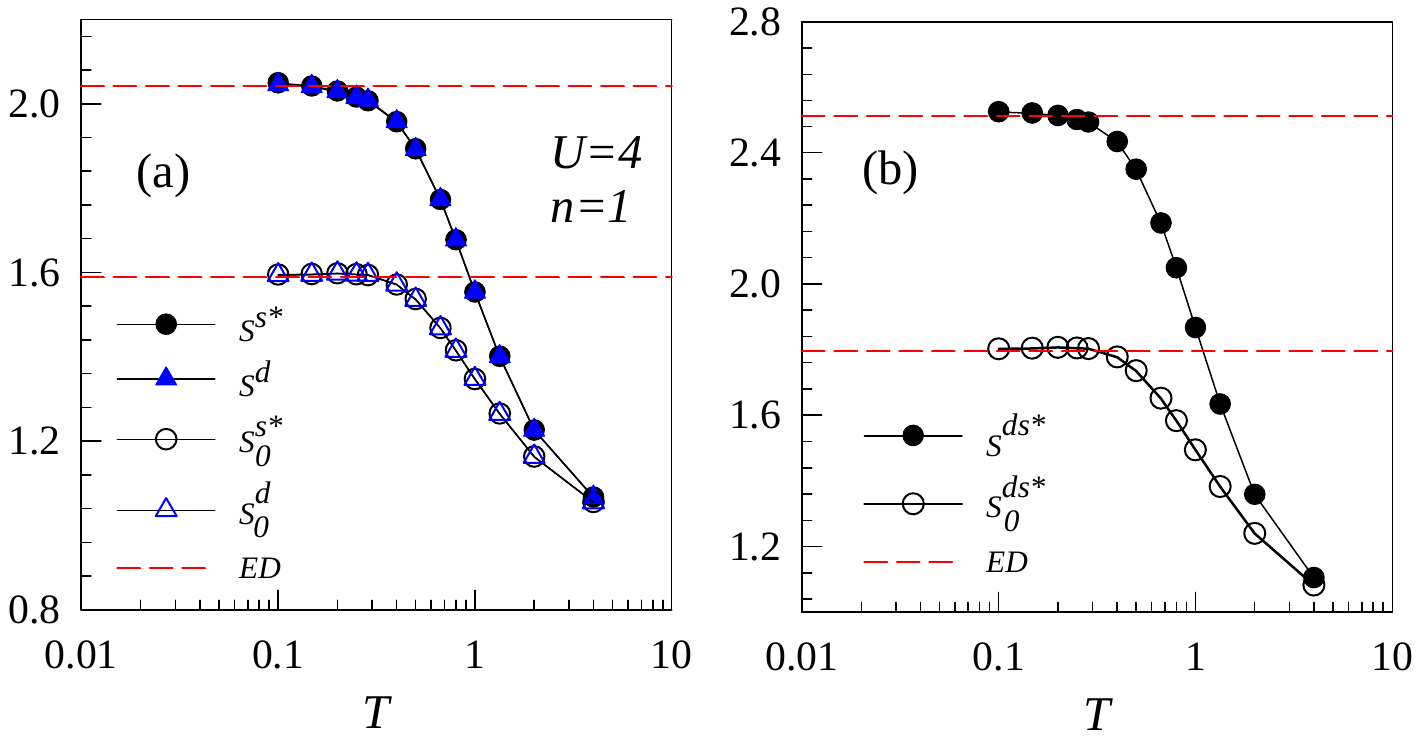}
\caption{The DQMC and ED results on $4\times 4$ lattice for $n=1$ and
$U=4$. The finite-temperature DQMC values tend to those of ED at zero
temperature.}
\label{figS4}
\end{figure}

To benchmark our DQMC simulations, we compare the DQMC results with
those from ED on small sizes. As shown in Fig.~\ref{figS4}, the finite-temperature
DQMC values for the pair structure factors of
all the symmetries precisely approach ED values at zero temperature.

\begin{figure}[htbp]
\centering \includegraphics[width=9cm]{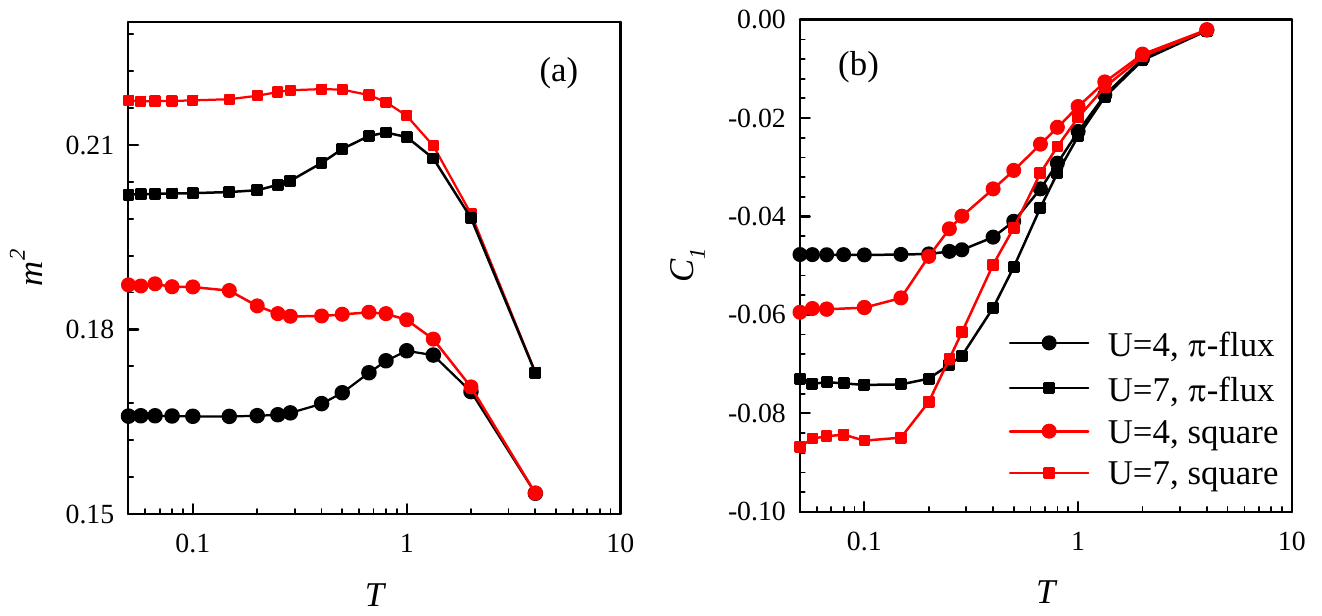} \caption{The comparision
of the local moment (a) and NN spin correlation (b) for fermions with
linear and quadratic dispersions. The results are extrapolated to the
continuous imaginary time limit using two separate
simulations with $\Delta \tau=\frac{1}{16}$ and $\Delta
\tau=\frac{1}{12}$. }
\label{figS5}
\end{figure}

\section{Effect of Flux on Local Magnetic Correlations}
Fig.~\ref{figS5} displays
the local moment $m^2$ and NN spin-spin correlation
function.  $m^2$ is the zero separation (${\bf l}=0$) value of
$C({\bf l})=\langle \frac{1}{2} (n_{{\bf
j+l}\uparrow}-n_{{\bf j+l}\downarrow})\frac{1}{2} (n_{{\bf j}\uparrow}-n_{{\bf
j}\downarrow})\rangle$
and reflects the degree of local charge
fluctuations (double occupancy). $C({\bf l})$ is rotationally invariant
and in our simulations we average over all three directions to provide
an improved estimator in DQMC simulations. As
shown in Fig.~S5(a), $m^2$ increases as $U$ is increased.
Although the two cases $\phi=0$ and $\phi=\pm \pi$,
have nearly the same $m^2$ at high
temperatures, this agreement
breaks down at $T/t \lesssim 1$: Dirac fermions have
smaller local moments at low
temperatures compared to fermions with quadratic dispersion. For the NN
spin correlation, at high temperatures  the
$\pi-$flux phase has bigger spin correlations, but there is a
crossover so that at low $T$ the $\phi=0$ lattice has larger
$C_1 = C\big({\bf l}=(1,0)\big)$.

\section{Divergence of the $ds^*$-wave pairing susceptibility}

\begin{figure}[htbp]
\centering \includegraphics[width=9cm]{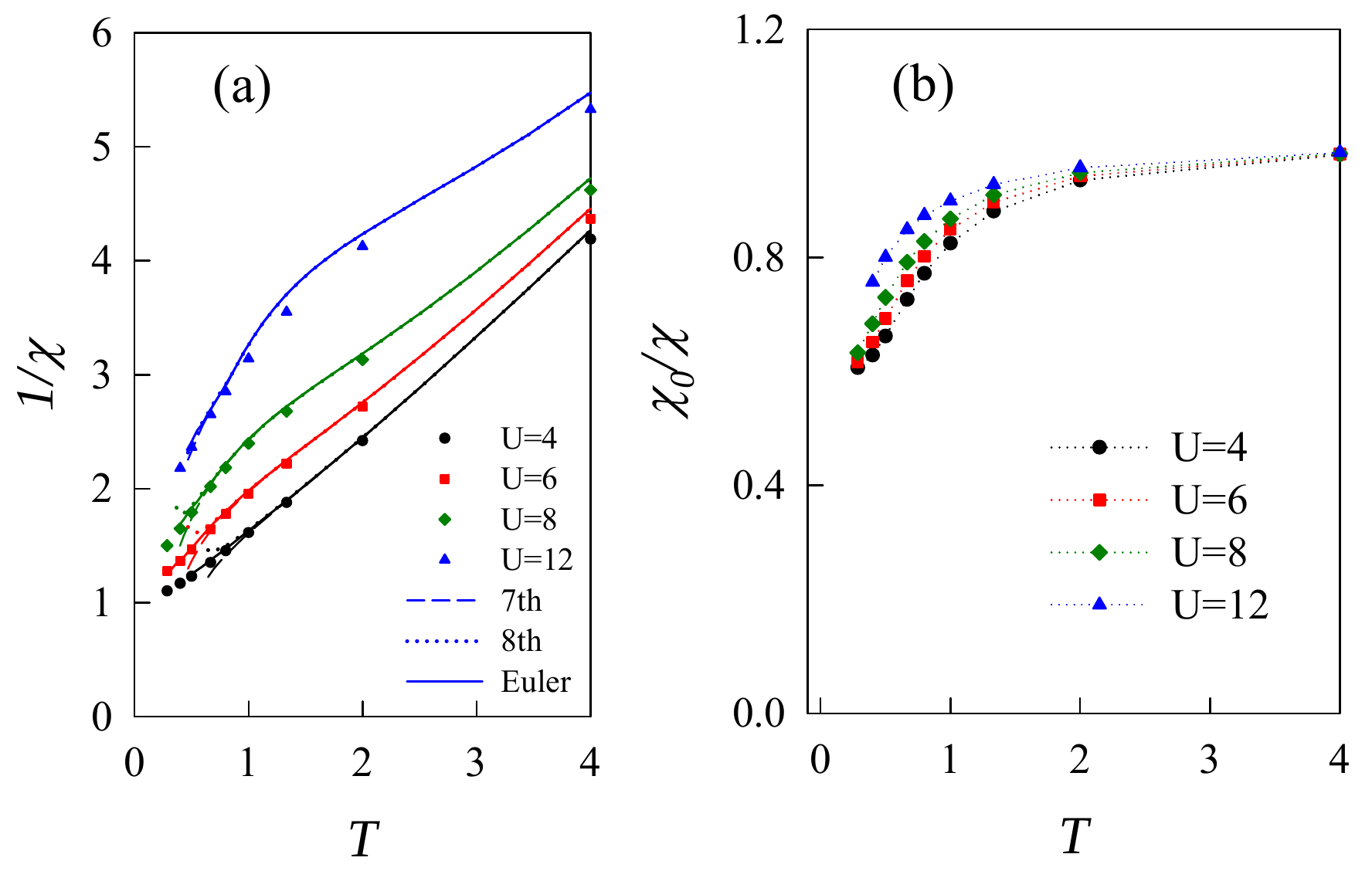} \caption{(a) The inverse of the $ds^*$-wave pairing susceptibility as a function of the temperature at $n=0.9$ for different values of $U$. (b) The inverse of the $ds^*$-wave pairing susceptibility divided by the local uncorrelated susceptibility at $r=0$.}
\label{figS6}
\end{figure}

At the superconducting transition temperature, the pairing susceptibility is expected to be divergent. Fig.~\ref{figS6} plots $1/\chi$ as a function of the temperature at $n=0.9$. The divergence of $\chi$, especially at small $U$, is not completely compelling. However as $U$ increases, the curves bend downward with growing slope and show an increasing tendency to cross zero at finite temperatures. To compare values of the
susceptibilities for different $U$ on a more equal footing, we divide the $ds^*$-wave pairing susceptibility by the local uncorrelated susceptibility at $r=0$. The scaled susceptibility dives more rapidly.  Due to the small density at the Fermi surface for the situation we considered, it is expected that superconductivity may happen at low temperature, which is beyond the current capabilities of the DQMC and NLCE  methods.

\section{NLCE Resummation}
 Similar to the Pade approximations widely used in high-temperature series expansions, in the NLCE, one can take advantage of numerical resummation techniques, such as the Euler or Wynn methods~\cite{nlce4,checkerboard}, to extend the region of convergence to lower temperatures. Here we use the Euler resummation for the last five terms in the series. In this method, the original sum is replaced by
\begin{equation}
S_1+S_2+S_3+\sum_{l=0}^{4}\frac{(-1)^l}{2^{l+1}}\Delta^{l}u_{4},
\end{equation}
where $S_n$ is the $n$th term in the series, $u_n=(-1)^nS_n$, and $\Delta$ is defined as the forward differencing operator
\begin{eqnarray}
\Delta^0 u_{n} &=& u_{n}, \nonumber \\
\Delta^1 u_{n} &=& u_{n+1} - u_{n}, \nonumber \\
\Delta^2 u_{n} &=& u_{n+2} - 2u_{n+1} + u_{n}, \nonumber \\
\Delta^3 u_{n} &=& u_{n+3} - 3u_{n+2} + 3u_{n+1} - u_{n}, \\
\vdots \nonumber
\end{eqnarray}


\begin{thebibliography}{100}
\bibitem{note1} Here we only consider conventional pairings on square lattice. Though there are other possible ones (such as time-reversal breaking $d+is^*$ et.al.), we find that $ds^*$-wave is the ground state of all allowed pairing symmetries. In our calculations, the pairing structure factor is formulated as $S=F^T M F$, with $M$ the paring matrix and $F$ the pairing wave.
The eigenvector corresponding to the largest eigenvalue of $M$ is $d^*s$-wave.
\end{thebibliography}
\end{document}